\documentclass[twocolumn,showpacs,preprintnumbers,amsmath,amssymb,superscriptaddress,aps]{revtex4}

\usepackage{graphicx}
\usepackage{dcolumn}
\usepackage{bm}

\begin{document}


\title{Light scattering observations of spin reversal excitations in the fractional quantum Hall regime }

\author{Irene Dujovne}
\affiliation{Department of Applied Physics and Applied Math,
Columbia University, New York, NY  10027} \affiliation{Bell Labs,
Lucent Technologies, Murray Hill, NJ 07974}

\author{C.F. Hirjibehedin}

\affiliation{Department of Physics, Columbia University, New York,
NY  10027} \affiliation{Bell Labs, Lucent Technologies, Murray
Hill, NJ 07974}

\author{A. Pinczuk}
\affiliation{Department of Applied Physics and Applied Math,
Columbia University, New York, NY  10027} \affiliation{Bell Labs,
Lucent Technologies, Murray Hill, NJ 07974}
\affiliation{Department of Physics, Columbia University, New York,
NY  10027}
\author{Moonsoo Kang}
\affiliation{Physics Department, Washington State University, WA
99164-2814.}

\author{B.S. Dennis}
\affiliation{Bell Labs, Lucent Technologies, Murray Hill, NJ
07974}

\author{L.N. Pfeiffer}
\affiliation{Bell Labs, Lucent Technologies, Murray Hill, NJ
07974}

\author{K.W. West}
\affiliation{Bell Labs, Lucent Technologies, Murray Hill, NJ
07974}

\date{\today}

\begin{abstract}
Resonant inelastic light scattering experiments access the low
lying excitations of electron liquids in the fractional quantum
Hall regime in the range $2/5 \geq \nu \geq 1/3$.  Modes
associated with changes in the charge and spin degrees of freedom
are measured. Spectra of spin reversed excitations at filling
factor $\nu \gtrsim 1/3$ and at $\nu \lesssim 2/5$ identify a
structure of lowest spin-split Landau levels of composite fermions
that is similar to that of electrons. Observations of spin wave
excitations enable determinations of energies required to reverse
spin. The spin reversal energies obtained from the spectra
illustrate the significant residual interactions of composite
fermions. At $\nu = 1/3$ energies of spin reversal modes are
larger but relatively close to spin conserving excitations that
are linked to activated transport. Predictions of composite
fermion theory are in good quantitative agreement with
experimental results.
\end{abstract}

\maketitle

The fractional quantum Hall effect (FQHE) is an electron
condensation phenomenon that occurs at low temperatures  when
two-dimensional electron systems of very low disorder are exposed
to high magnetic fields. The FQH states are archetypes of quantum
fluids that  emerge in low dimensional electron systems due to the
impact of fundamental interactions. In the FQHE the 2D electron
system becomes incompressible at certain values of the Landau
level filling factor $\nu=n h c / e B$, where $n$ is the electron
density and B is the perpendicular magnetic field. In the filling
factor range $1\geq \nu \geq 1/3$ the major sequence of the FQHE
occurs at `magic' filling factors $\nu = p/(2p\pm 1)$, where $p$
is an integer. The composite fermion (CF) framework interprets the
sequence by attaching two vortices of the many body wavefunction
to each electron \cite{jain,heinonen}. Chern-Simons gauge fields
incorporate electron interactions so that CF's experience
effective magnetic fields $B^*= B - B_{1/2} = \pm B/(2p\pm 1)$,
where $B_{1/2}$ is the perpendicular magnetic field at $\nu= 1/2$
\cite{hlr,kalm,lopez91}. Composite fermion quasiparticles have
spin-split energy levels characteristic of charged fermions with
spin 1/2 moving in the effective magnetic field $B^*$. The levels
resemble spin-split Landau levels of electrons. The number $p$
thus becomes the CF Landau level filling factor and in FQHE states
at $\nu = p/(2p\pm 1)$ there are $p$ fully occupied levels of
composite fermions.
\par
\begin{figure*}
\includegraphics{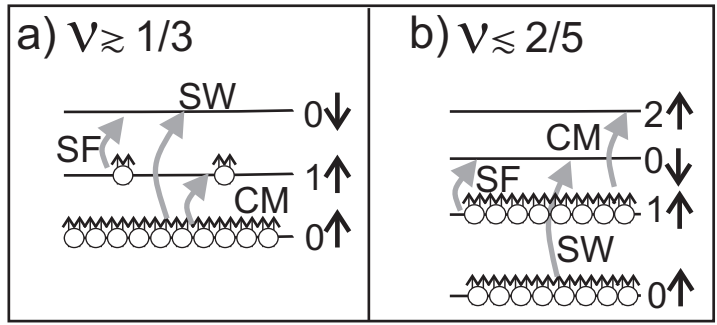}
\caption{\label{levels} Structure of spin-split Landau levels of
CFs that interpret the low lying excitations observed in this
work. (a) $\nu \gtrsim 1/3$. (b) $\nu \lesssim 2/5$. Landau levels
are labelled by quantum numbers and arrows that indicate
orientation of spin. The lowest SF, SW and CM transitions are
depicted. Composite fermions are shown as small circles with two
arrows that represent the two vortices attached to each electron.
}
\end{figure*}

Structures of spin-split CF levels are shown schematically in Fig.
\ref{levels} for $\nu \gtrsim 1/3$ and $\nu \lesssim 2/5$. The
spacing between sequential CF levels with same spin is represented
as a cyclotron frequency
\cite{hlr,fradkin,du93,simon,park,murthy99,mandal,aoki01},
$\omega_c=e B^*/c m_{CF}$, where $m_{CF}$ is a CF effective mass.
Figure \ref{levels} also presents the transitions of composite
fermions near filling factors 1/3 and 2/5. The charge mode (CM)
transitions are spin-conserving. The spin wave (SW) and spin flip
(SF) transitions involve a spin-reversal. The CM transition
energies, however, could be different from CF level spacings. The
difference is due to the changes in self-interaction energies that
may occur when quasiparticles and quasiholes are created
\cite{park}.
\par
Neutral pair excitations of quasiparticles above FQHE ground
states are constructed from transitions such as those shown in
Fig. \ref{levels}. The excitations are described as modes that
have characteristic energy vs. wave vector dispersions $\omega
(q)$\cite{kallin,haldane,girvin}. The separation between the
neutral quasiparticle-quasihole pairs is $x_o = ql_o^{2}$, where
$l_o = (\hbar c/eB)^{1/2}$ is the magnetic length. The Coulomb
energy $E_c=e^{2}/\epsilon l_{o}$, where $\epsilon$ is the
dielectric constant of the host semiconductor ($\epsilon = 12.6$
in GaAs) defines the energy scale for the excitations. Salient
features of the dispersions are at critical points. The critical
points occur at $q \rightarrow 0$, at the $q \rightarrow \infty$
limit and at roton minima at finite wavevector. These roton
minima, known as magnetorotons, are due to excitonic binding
interactions in neutral quasiparticle-quasihole pairs with $q
\simeq 1/l_{o}$. The mode at large wavevector
($q\rightarrow\infty$) represents a non-interacting
quasiparticle-quasihole pair at large separation \cite{hlr}.
\par
Composite fermion quasiparticles are expected to experience
interactions that are much weaker than those among electrons. This
occurs because Chern-Simons gauge fields linked to vortex
attachment in forming CF quasiparticles incorporate significant
Coulomb interactions. The energy differences between critical
points, such as that between the magnetoroton and the large wave
vector ($q \rightarrow \infty$) CM excitations, being
manifestations of residual interactions, provide experimental
venues to test key predictions of the composite fermion approach.
\par
\par
The critical points of dispersions of spin and charge collective
modes of the FQH liquid are accessed in resonant inelastic light
scattering experiments
\cite{pinc93,pinc95,davies97,perspectives,davies99,moonsoo00}.
Results at $\nu = 1/3$ have determined the energies of rotons at
$q\sim 1/l_o$, and of large wave vector ($q\rightarrow\infty$)
excitations of the spin conserving CM transitions
\cite{moonsoo01}. The observed energy splitting between rotons at
$\Delta_{R}$ and $q\rightarrow\infty$ modes at $\Delta_{\infty}$,
by light scattering measurements, is direct evidence of the
strength of residual CF interactions. CF evaluations of
dispersions of quasiparticle excitations are in good quantitative
agreement with light scattering results \cite{moonsoo01,scar00}.
\par
In more recent light scattering experiments quasiparticle
excitations in the spin and charge degrees of freedom have been
observed in the full range $2/5\geq \nu \geq 1/3$
\cite{dujovne,dujovne03}. The measured excitations are interpreted
with transitions CM, SW and SF. The results suggest that CF
quasiparticles have the well defined structure of spin-split CF
Landau levels shown in Fig \ref{levels}. The dependence of the
measured mode energies on filling factor in the range $2/5\geq \nu
\geq 1/3$ indicate significant residual interactions among the
quasiparticles.
\par
We report here inelastic light scattering studies of spin
excitations of the electron liquid at filling factors $1/3 \leq
\nu \leq 2/5$. Two distinct types of spin modes are considered:
spin waves and spin flip excitations based on the correspondingly
labelled transitions shown in Fig.\ref{levels}. In spin waves
there is only spin reversal, while in spin flip modes spin and CF
Landau level quantum number change simultaneously.
\par
The light scattering measurements of spin excitations yield direct
determinations of spin-reversal energies of quasiparticles of the
FQH liquids. The results reveal large spin reversal energies that
are due to residual quasiparticle interactions
\cite{tapash,longo,aoki,lopez95,mandalactivation}. In conjunction
with determinations of CM transitions the light scattering
measurements of spin excitations represent unique experimental
tests of composite fermion theory. These large residual CF
interactions are possibly linked to those that manifest in
condensation of composite fermions into higher order liquid states
\cite{pan03,scar02,izab02}.
\par
At filling factor 1/3 we identify the large wavevector limit
($q\rightarrow\infty$) of the spin wave energy $E_Z^*$ at
\begin{equation}
E_{Z}^{*}=E_{Z}+E^{\uparrow\downarrow}\label{ez}
\end{equation}
where $E^{\uparrow\downarrow}$ represents the spin reversal energy
due to interactions among quasiparticles
\cite{tapash,longo,aoki,lopez95,mandalactivation}. The bare Zeeman
energy is $E_Z=g\mu_B B_T$, where $B_T$ is the total magnetic
field, $\mu_B$ is the Bohr magneton and $g \sim 0.44$ is the
absolute value of the Lande factor of GaAs. A peak at $E_{Z}$ that
occurs in the light scattering spectra is interpreted as the
$q\rightarrow 0$ limit of the SW mode (Larmor's theorem)
\cite{kallin}. The light scattering observations of SW modes at
large wave vector ($q\rightarrow \infty$) at $E_{Z}^{*}$ reported
here allow the direct determination of $E^{\uparrow\downarrow}$.
\par
A determination of $E^{\uparrow\downarrow}$ was reported in Ref.
\cite{dujovne} from the energy of $|1\uparrow> \rightarrow
|0\downarrow>$ transitions that occur when the second CF Landau
level begins to populate at $\nu \gtrsim 1/3$. In Ref.
\cite{dujovne} the simultaneous measurements of $E_{Z}$, the CM
collective mode at $\Delta_{\infty}$ and the newly discovered SF
mode enabled a preliminary determination of the many-body spin
reversal energy. Both $\Delta_{\infty}$ and
$E^{\uparrow\downarrow}$ are linear in the Coulomb energy $E_c$.
The value of $E^{\uparrow\downarrow}\cong$ 0.054$E_c$ has been
reported. The alternate determination of $E^{\uparrow\downarrow}$
from large wave vector ($q \rightarrow\infty$) SW modes reported
here is in excellent agreement with the previous result.
\par
In the present study we find that $\Delta_{\infty}$ is smaller but
close to the determined value of $E_{Z}^{*}$. $\Delta_{\infty}$,
the energy of a widely separated spin conserving quasiparticle
quasihole pair, is linked to the activation energy that determines
the temperature dependence of longitudinal magnetotransport. The
result showing that $\Delta_{\infty}$ and $E_{Z}^{*}$ have similar
energy implies that spin excitations could play a role in
temperature activated processes and explain some of the
discrepancies between experimental and calculated activation
energies \cite{du93,mandalactivation}.
\par
\begin{figure*}
\includegraphics{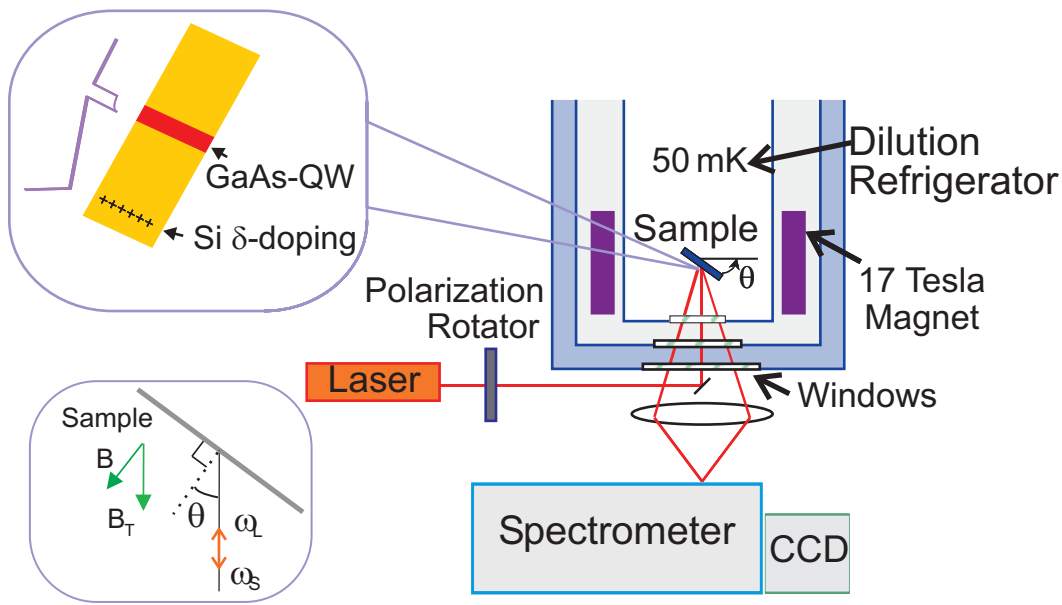}
\caption{\label{fridge+sample} Schematic description of light
scattering experiments at millikelvin temperatures. The top inset
depicts quantum layers and the bottom of conduction. The lower
inset shows the backscattering arrangement for the sample and the
orientation of the magnetic field.}
\end{figure*}
\par
In the composite fermion framework the interplay between spin
reversal energies and level spacings dictates the spin
polarization of the liquid states. In this competition between the
two energies the small Zeeman energy can play a pivotal role. For
example, states such as 2/5, with two CF Landau levels fully
populated (as seen in Fig. \ref{levels}), are spin-polarized
because at relatively large fields $E_{Z}$ is larger than the
difference between the level spacing and the spin reversal energy
due to interactions. The interaction term of the spin reversal
energy, $E^{\uparrow\downarrow}$, depends on the spin polarization
of the FQH states. Light scattering determinations of spin
reversal energies could offer venues to probe spin polarizations
of states with $\nu > 1/3$.

\par
The resonant inelastic light scattering experiments are carried
out with photon energies that overlap fundamental optical
transitions of the semiconductor quantum structure that hosts the
2D system. Conservation of energy in the inelastic scattering
processes is expressed as $\omega(q)=\pm (\omega_L-\omega_S)$,
with $\pm$ corresponding to the stokes and anti-stokes processes
respectively. The in-plane inelastic scattering wavevector,
$\mathbf{k_{\parallel}}$, is such that
$|\mathbf{k_{\parallel}}|=k_{\parallel}=(k^L-k^S)$sin$\theta \sim
(2 \omega_{L}/c) \sin\theta$, where L and S refer to incident
(laser) and scattered wavevectors of light and $\theta$ is the
tilt angle. The backscattering configuration shown in Fig
\ref{fridge+sample} offers access to in-plane wavevectors
$k_{\parallel} \thicksim 10^5$~cm$^{-1}$ and $k_{\parallel} l_{o}
\lesssim 0.1$. Conservation of momentum for systems with
translational invariance is equivalent to conservation of
wavevector. In 2D systems this converts to
$\mathbf{q}=\mathbf{k_{\parallel}}$. Wavevector conservation
breaks down with the loss of full translation symmetry in the
presence of weak residual disorder that occurs even in systems of
high perfection. In this case the light scattering spectra will
display the contributions from the critical points in the density
of states \cite{ pinc88,pinc95,davies97,marmorkos}.
\par
The high quality 2D electron system in each sample resides in
single GaAs quantum well (SQW) of typical width $d=330$~\AA\/. We
present results from samples with density $n = 5.6\times
10^{10}$~cm$^{-2}$ and $7.6\times 10^{10}$~cm$^{-2}$. The electron
mobilities are very high, reaching a value
$\mu\gtrsim7\times10^{6}cm^{2}/Vsec$ at T $ \cong 300mK$. The
samples were mounted on the cold finger of a
${}^{3}\text{He}/{}^{4}\text{He}$ dilution refrigerator that is
inserted in the cold bore of a superconducting magnet with windows
for optical access (see Fig. \ref{fridge+sample}). Cold finger
temperatures are variable and as low as 45~mK. The resonant
inelastic light spectra were excited with a diode laser with
photon energies $\omega_{L}$ close to the fundamental optical gap
of the GaAs SQW. The power density was kept below
$10^{-4}$~W/cm$^2$. The spectra were acquired  with a double
Czerny-Turner spectrometer operating in additive mode and a CCD
camera with 15 $\mu$m pixels. The combined resolution with a 30
$\mu$m slit is 0.016 meV.
\par
We measured depolarized spectra with orthogonal incident and
scattered light polarizations and polarized spectra in which
polarizations are parallel. Excitation modes with changes in the
spin degree of freedom tend to be stronger in depolarized spectra,
while modes in the charge degree of freedom tend to be stronger in
polarized spectra. A backscattering geometry shown in the inset to
Fig. \ref{fridge+sample} was used with an angle $\theta \sim
30^o$. The perpendicular component of magnetic field is
$B=B_{T}\cos\theta$, where $B_{T}$ is the total field.
\par
\begin{figure}
\includegraphics{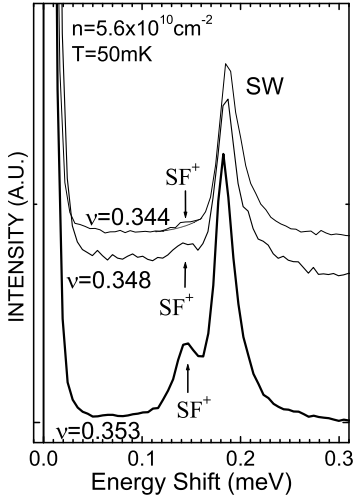}
\caption{\label{near1/3} Depolarized inelastic light scattering
spectra at filling factors $\nu \gtrsim 1/3$. Vertical arrows show
the peak assigned to the spin-flip mode}
\end{figure}
\par
Figure \ref{near1/3} displays depolarized spectra at filling
factors $\nu \gtrsim 1/3$. The spectra show the long wavelength SW
at $E_Z$ and a weaker narrow peak, labelled SF$^+$, that occurs at
lower energies. The intensity of the new peak depends strongly on
filling factor. It is absent for $\nu = 1/3$ and its intensity
increases as the field is decreased towards higher filling
factors, indicating that this mode is related to the population of
the second Landau level $|1\uparrow>$. On this basis this mode was
assigned to the SF transition $|1\uparrow> \rightarrow
|0\downarrow>$  shown in Fig. \ref{levels} \cite{dujovne}.
\par
\begin{figure}
\includegraphics{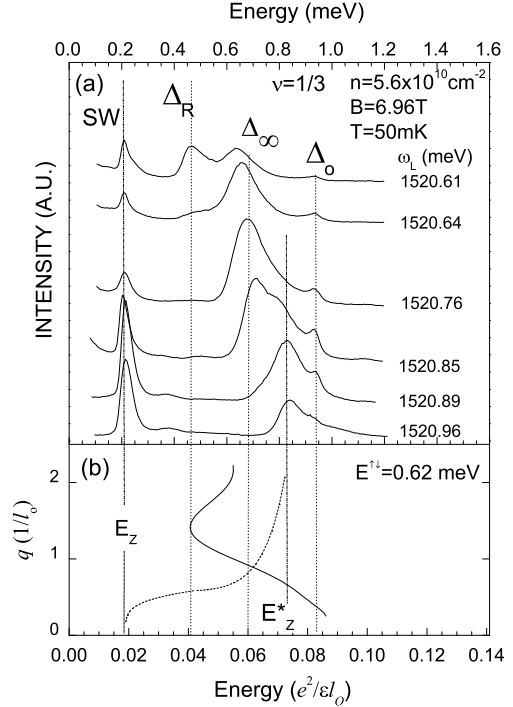}
\caption{\label{fig-1/3} (a) Resonant inelastic light scattering
spectra at $\nu$=1/3.  The incident photon energies are indicated
in meV. The mode labelled SW is the $q=0$ spin wave excitation at
E$_z$ . $\Delta_R$ is the CM roton and $\Delta_0$ is the CM mode
at $q=0$. The large wavevector limit ($q \rightarrow \infty$) of
the CM mode is at $\Delta_{\infty}$. $E_{Z}^{*}$ is the large
wavevector limit ($q \rightarrow \infty$) of the spin wave mode.
(b) The thick line is the dispersion of the CM excitation (after
Scarola et al. \cite{scar00}). The dotted line is a schematic
representation for the spin wave dispersion.}
\end{figure}

Figure \ref{fig-1/3}(a) displays spectra taken at $\nu = 1/3$.
Four of the excitations shown in Fig. \ref{fig-1/3}(a) have been
assigned in Ref. \cite{moonsoo01}. SW is the long wavelength
($q\rightarrow 0$) spin wave mode at E$_z$. The other three are CM
modes: $\Delta_R$ is the magnetoroton, $\Delta_0$ is the $q
\rightarrow 0$ mode, and $\Delta_{\infty}$ the large wavevector
mode. The assignments are consistent with the calculated
dispersions of CM modes shown in Fig.
\ref{fig-1/3}(b)\cite{scar00,moonsoo01}. There is a fifth mode
that has not been previously reported. It appears at an energy of
0.83meV, between $\Delta_0$ and $\Delta_{\infty}$. It is natural
to assign this mode to the critical point of spin wave excitations
that occurs at large wavevectors. Its energy is indicated as
$E_{Z}^{*}$ (see Eq. \ref{ez}). This assignment enabled us to
construct the schematic rendition of the spin wave dispersion
shown by the dotted line curve in Fig \ref{fig-1/3}(b).
\par
The spin flip excitation labelled SF$^+$ in Fig. \ref{near1/3}
occurs at $\nu \gtrsim 1/3$, when there is only very low
population of the $|1\uparrow>$ level. The mode is represented by
the SF transitions shown in Fig. \ref{levels}. Its energy is
approximated as \cite{dujovne}
\begin{equation}
E_{SF}=E_Z+E^{\uparrow\downarrow}-\Delta_{\infty} \label{esf}
\end{equation}
The value $\Delta_{\infty} = 0.059 E_c$ was obtained from the CM
mode at large $q$ shown in Fig. \ref{fig-1/3}(a).  The bare Zeeman
energy was obtained from spectra such as those in Figs.
\ref{near1/3} and \ref{fig-1/3}(a), where $E_{Z}$ is given by the
position of the peaks labeled SW. With these determinations it was
found that E$^{\uparrow\downarrow} \cong 0.054 E_c$
\cite{dujovne}.
\par
The measurement of the total spin reversal energy $E_{Z}^{*}$ from
the spin wave mode at $q\rightarrow\infty$, as shown in Fig.
\ref{fig-1/3}, provides a second determination of the many-body
term of the spin reversal energy $E^{\uparrow\downarrow}$. From
$E_{Z}^{*} =$ 0.83meV and $E_{Z}$ = 0.21meV we obtain
$E^{\uparrow\downarrow}=$0.62 meV$= 0.053E_c$, identical within
the experimental uncertainty to the value of
$E^{\uparrow\downarrow}$ obtained from measurements of $E_{SF}$.
This consistency between the two independently obtained values
testifies to the significant reliability of the determinations of
$E_{Z}^{*}$ and $E^{\uparrow\downarrow}$ from light scattering
spectra at $\nu\gtrsim 1/3$.
\par
\par
\begin{figure*}
\includegraphics{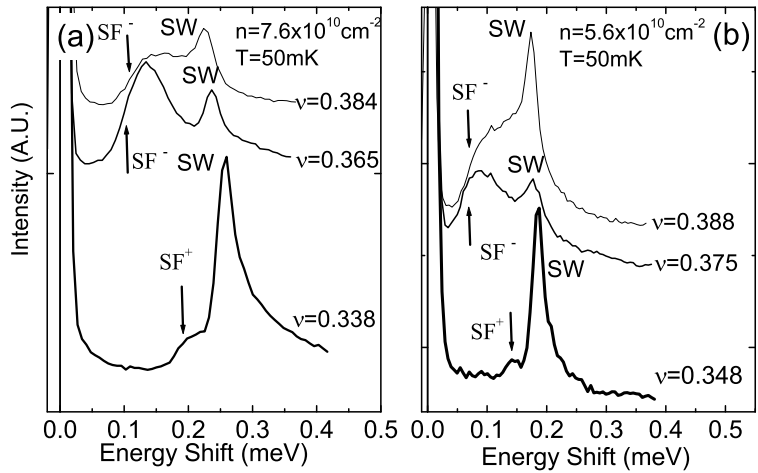}
\caption{\label{fig-10-20} Inelastic light scattering measurements
at different filling factors $1/3 \geq \nu \geq 2/5$ for two
densities (a) $7.6 \times10^{10}$~cm$^{-2}$ and (b) $5.6 \times
10^{10}$~cm$^{-2}$ . Vertical arrows shows the peaks assigned to
the SF modes (SF$^-$ and SF$^+$)}
\end{figure*}
\par
Figure \ref{fig-10-20} shows spectra of low lying spin excitation
modes at filling factors in the range $2/5\geq \nu \geq 1/3$. The
results in panel (a) are from the sample with density $7.6 \times
10^{10}$ cm$^{-2}$. The spectra in panel (b), from the sample with
density $5.6 \times10^{10}$ cm$^{-2}$, are those reported in Ref.
\cite{dujovne}. Although there is considerable difference in
density, both samples show a similar behavior. The figure reveals
that an excitation derived from the SF roton at $\nu=2/5$
\cite{mandal,dujovne}, the mode labelled SF$^-$, shifts to lower
energies as the second level starts to depopulate for $\nu\lesssim
2/5$, and disappears for filling factors in the middle of the
range $2/5\geq \nu \geq 1/3$. At larger magnetic fields, $\nu
\gtrsim 1/3$, a narrow peak (SF$^+$) appears below the SW. This
peak, also displayed in Fig. \ref{near1/3}, disappears as the
state with $\nu=1/3$ is approached. The results at $\nu \gtrsim
1/3$ in Fig. \ref{fig-10-20}(a) enable determinations of
$E^{\uparrow\downarrow}$ with Eq. \ref{esf}. We find a square root
dependence on perpendicular field given by $E^{\uparrow\downarrow}
\cong 0.054E_c$ at both densities.
\par
\begin{figure}
\includegraphics{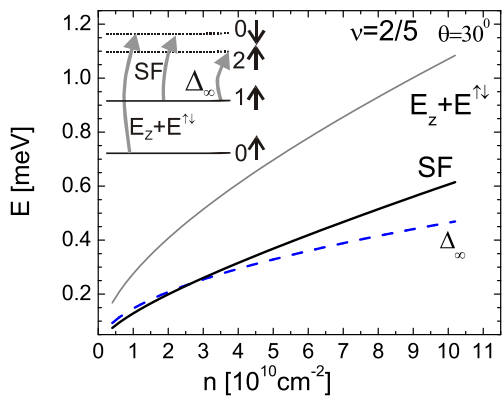}
\caption{\label{density-2/5} Evaluations of the energies of the
large wavevector limits of the spin wave at
$E_{Z}+E^{\uparrow\downarrow}$, of the CM mode at $\Delta_\infty$
and of the spin flip (SF) modes as function of density for
$\nu=2/5$ and $\theta = 30^o$. The inset shows the lowest CF
levels. Fully populated levels are depicted as full lines. Empty
levels are shown as dotted lines. }
\end{figure}
\par
The values of $\Delta_{\infty}$, $E^{\uparrow\downarrow}$ and
$E_{Z}$ determined at $\nu = 1/3$ are employed to estimate the
density dependence of the low-lying spin and charge excitations at
$\nu=1/3$ and $\nu=2/5$. The linear dependence of
$\Delta_{\infty}$ and $E^{\uparrow\downarrow}$ on $E_c$ translates
into a square root dependence on density. For each value of
density we assume that the CM energy at large wavevector for
$\nu=2/5$ is given by the simple scaling
$\Delta_{\infty}(2/5)=\Delta_{\infty}/2$.
\par
Figure \ref{density-2/5} displays evaluations of the energies of
low-lying spin and charge excitations as function of density for
an angle $\theta = 30^o$ at filling factor $\nu=2/5$ ($p=2$). The
figure shows the large wave vector spin wave energy
$E_{Z}+E^{\uparrow\downarrow}$, $\Delta_{\infty}(2/5)$ and the
energy of the large wave vector spin-flip mode (SF). The values
obtained for the SF energy are in excellent agreement to the
composite fermion evaluations by Mandal and Jain at
\cite{mandal,mandalactivation}. On the other hand our values
$\Delta_{\infty}(2/5)$ are lower than those in Ref.
\cite{mandalactivation}.
\par
One of the intriguing issues to consider further is that of the
relation between SF modes and the spin polarization of the 2/5
state. A loss of spin polarization in the 2/5 state was reported
in transport experiments with tunable Zeeman energy and in
luminescence determinations as function of density
\cite{wkang,kukushkin}. Recently a roton in the dispersion of SF
excitations was discovered \cite{mandal,dujovne}. It is
conceivable that the loss of polarization is related to a roton
instability of SF excitations that occurs when $E_Z$ collapses
either at small densities or small $g$ factor.

\par
\begin{figure}
\includegraphics{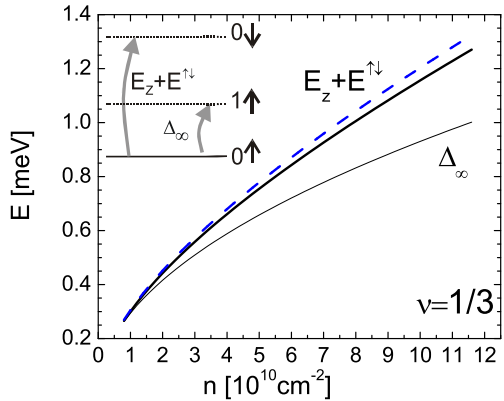}
\caption{\label{density} Evaluations at $\nu=1/3$ of large
wavevector limit of the SW (thick line), spin reversal energy and
$\Delta_\infty$ (thin line) as a function of density. The large
wavevector limit of the spin wave is calculated at 30$^o$ (dashed
line) and 0$^o$(solid line). The inset shows the lowest CF levels.
Fully populated levels are shown as full lines. Empty levels are
represented as dotted lines.}
\end{figure}
\par
Similar evaluations of the density dependence of the
$\Delta_{\infty}$, $E_{Z}^{*}$ and $E^{\uparrow\downarrow}$
energies evaluated at $\nu=1/3$ are presented in Fig.
\ref{density}. The finding that $\Delta_{\infty}$is relatively
close to $E_{Z}^{*}$ confirms that the bare Zeeman energy plays
the key role in setting up the hierarchy of the lowest CF levels
and excitations at $q \rightarrow \infty $. We recall that
$\Delta_{\infty}$ is considered to be the activation energy
measured in magnetotransport. The closeness of $\Delta_{\infty}$
and $E_{Z}^{*}$, shown Fig. \ref{density}, suggest that spin
reversal excitations represented by $E_{Z}^{*}$ could have a non
trivial impact on charge transport at $\nu = 1/3$.
\par
The results for $\Delta_{\infty}$ in Fig. \ref{density} have
played a key role in the determination of of spin reversal
energies \cite{dujovne}. $\Delta_{\infty}$ can be represented by a
cyclotron frequency with CF cyclotron mass $m_{CF}$ that is a
function of $n^{1/2}$. For the two samples of this study we find
masses of 0.39m$_e$ and 0.45m$_e$ for the densities 5.6 and $7.6
\times 10^{10}$ cm$^{-2}$ respectively. At $n= 1.1 \times
10^{11}$cm$^{-2}$ the $n^{1/2}$ dependence of the cyclotron mass
yields a value $m_{CF} = 0.54$m$_e$, that is somewhat smaller, but
close, than determinations of $m_c$ from activated
magnetotransport \cite{du93}.
\par
The results presented in this paper show that light scattering
experiments offer direct access to key features of low-lying
excitations in the spin and charge degrees of freedom in the FQH
regime. Near 2/5 and 1/3 the spectra offer evidence of spin-split
CF Landau levels of charged fermions with spin 1/2. Spin-reversal
energies, associated with spin waves and spin flip excitations,
are readily determined from the spectra. Current evaluations of
spin-reversed quasiparticle excitations are in excellent
quantitative agreement with our results. The spin reversal
energies, typically about one fifth of that for electrons at $\nu
=1$ \cite{pinc92}, indicate significant residual CF interactions.
These residual interactions could interpret condensed states at
filling factors in the range 2/5 $>\nu>$ 1/3 with partial
population of CF Landau levels\cite{pan,izab,scar02}.
\par
Inelastic light scattering methods at low temperatures in the
100mK range and below have enormous potential to probe the impact
of fundamental interactions of two dimensional electron systems in
the FQH regimes. By giving unique access to low lying spin
reversal excitation light scattering experiments can study the
spin polarization of the liquid states. Determinations of
$\Delta_{\infty}(\nu)$ when compared with the corresponding
activation energies in magnetotransport will enable assessments of
mechanisms for charge transport. Further studies of low lying
modes could offer unprecedented insights into quantum phases of
the low-dimensional electron system.
\par
We are grateful to H. L. Stormer for discussions on this work.
This work was supported in part by the Nanoscale Science and
Engineering Initiative of the National Science Foundation under
NSF Award Number CHE-0117752 and by a research grant of the W. M.
Keck Foundation.


\begin{thebibliography}{40}
\expandafter\ifx\csname
natexlab\endcsname\relax\def\natexlab#1{#1}\fi
\expandafter\ifx\csname bibnamefont\endcsname\relax
  \def\bibnamefont#1{#1}\fi
\expandafter\ifx\csname bibfnamefont\endcsname\relax
  \def\bibfnamefont#1{#1}\fi
\expandafter\ifx\csname citenamefont\endcsname\relax
  \def\citenamefont#1{#1}\fi
\expandafter\ifx\csname url\endcsname\relax
  \def\url#1{\texttt{#1}}\fi
\expandafter\ifx\csname
urlprefix\endcsname\relax\def\urlprefix{URL }\fi
\providecommand{\bibinfo}[2]{#2}
\providecommand{\eprint}[2][]{\url{#2}}

\bibitem[{\citenamefont{Jain}(1989)}]{jain}
\bibinfo{author}{\bibfnamefont{J.~K.} \bibnamefont{Jain}},
  \bibinfo{journal}{Phys.\ Rev. \ Lett.} {\bibinfo{volume}{63}}
 (\bibinfo{year}{1989}) \bibinfo{pages}{199} .

\bibitem[{\citenamefont{Heinonen}(1998)}]{heinonen}
\bibinfo{editor}{\bibfnamefont{O.}~\bibnamefont{Heinonen}}, ed.,
  \emph{\bibinfo{title}{Composite fermions: a unified view of the quantum Hall
  regime}} (\bibinfo{publisher}{World Scientific}, \bibinfo{year}{1998}).

\bibitem[{\citenamefont{Halperin et~al.}(1993)\citenamefont{Halperin, Lee, and
  Read}}]{hlr}
\bibinfo{author}{\bibfnamefont{B.~I.} \bibnamefont{Halperin}},
  \bibinfo{author}{\bibfnamefont{P.~A.} \bibnamefont{Lee}}, \bibnamefont{and}
  \bibinfo{author}{\bibfnamefont{N.}~\bibnamefont{Read}},
  \bibinfo{journal}{Phys.\ Rev. \ B}
  {\bibinfo{volume}{47}} (\bibinfo{year}{1993})
  \bibinfo{pages}{7312} .

\bibitem[{\citenamefont{Kalmeyer and Zhang}(1992)}]{kalm}
\bibinfo{author}{\bibfnamefont{V.}~\bibnamefont{Kalmeyer}} \bibnamefont{and}
  \bibinfo{author}{\bibfnamefont{S.-C.} \bibnamefont{Zhang}},
  \bibinfo{journal}{Phys.\ Rev. \ B} {\bibinfo{volume}{46}}
  (\bibinfo{year}{1992})
  \bibinfo{pages}{9889}.

\bibitem[{\citenamefont{Lopez and Fradkin}(1991)}]{lopez91}
\bibinfo{author}{\bibfnamefont{A.}~\bibnamefont{Lopez}} \bibnamefont{and}
  \bibinfo{author}{\bibfnamefont{E.}~\bibnamefont{Fradkin}},
  \bibinfo{journal}{Phys.\ Rev. \ B} {\bibinfo{volume}{44}} (\bibinfo{year}{1991})
  \bibinfo{pages}{5246}.

\bibitem[{\citenamefont{Lopez and Fradkin}(1993)}]{fradkin}
\bibinfo{author}{\bibfnamefont{A.}~\bibnamefont{Lopez}} \bibnamefont{and}
  \bibinfo{author}{\bibfnamefont{E.}~\bibnamefont{Fradkin}},
  \bibinfo{journal}{Phys.\ Rev. \ B} {\bibinfo{volume}{47}} (\bibinfo{year}{1993})
  \bibinfo{pages}{7080}.

\bibitem[{\citenamefont{Du et~al.}(1993)\citenamefont{Du, Stormer, Tsui,
  Pfeiffer, and West}}]{du93}
\bibinfo{author}{\bibfnamefont{R.}~\bibnamefont{Du}},
  \bibinfo{author}{\bibfnamefont{H.}~\bibnamefont{Stormer}},
  \bibinfo{author}{\bibfnamefont{D.}~\bibnamefont{Tsui}},
  \bibinfo{author}{\bibfnamefont{L.}~\bibnamefont{Pfeiffer}}, \bibnamefont{and}
  \bibinfo{author}{\bibfnamefont{K.}~\bibnamefont{West}},
  \bibinfo{journal}{Phys.\ Rev. \ Lett.} {\bibinfo{volume}{70}} (\bibinfo{year}{1993})
  \bibinfo{pages}{2944}.

\bibitem[{\citenamefont{Simon and Halperin}(1993)}]{simon}
\bibinfo{author}{\bibfnamefont{S.~H.} \bibnamefont{Simon}} \bibnamefont{and}
  \bibinfo{author}{\bibfnamefont{B.~I.} \bibnamefont{Halperin}},
  \bibinfo{journal}{Phys.\ Rev. \ B} {\bibinfo{volume}{48}} (\bibinfo{year}{1993})
  \bibinfo{pages}{17368}.

\bibitem[{\citenamefont{Park and Jain}(1998)}]{park}
\bibinfo{author}{\bibfnamefont{K.}~\bibnamefont{Park}} \bibnamefont{and}
  \bibinfo{author}{\bibfnamefont{J.~K.} \bibnamefont{Jain}},
  \bibinfo{journal}{Phys.\ Rev. \ Lett.} {\bibinfo{volume}{80}} (\bibinfo{year}{1998})
  \bibinfo{pages}{4237}; \bibinfo{journal}{Solid State Comm.} {\bibinfo{volume}{119}} (\bibinfo{year}{2001})
  \bibinfo{pages}{291}.

\bibitem[{\citenamefont{Murthy}(1999)}]{murthy99}
\bibinfo{author}{\bibfnamefont{G.}~\bibnamefont{Murthy}},
  \bibinfo{journal}{Phys.\ Rev. B} {\bibinfo{volume}{60}}(\bibinfo{year}{1999})
  \bibinfo{pages}{13702} .

\bibitem[{\citenamefont{Mandal and Jain}(2001{\natexlab{a}})}]{mandal}
\bibinfo{author}{\bibfnamefont{S.~S.} \bibnamefont{Mandal}} \bibnamefont{and}
  \bibinfo{author}{\bibfnamefont{J.~K.} \bibnamefont{Jain}},
  \bibinfo{journal}{Phys.\ Rev. B} {\bibinfo{volume}{63}} (\bibinfo{year}{2001}{\natexlab{a}})
  \bibinfo{pages}{201310}.

\bibitem[{\citenamefont{Onoda et~al.}(2001)\citenamefont{Onoda, Mizusaki, and
  Aoki}}]{aoki01}
\bibinfo{author}{\bibfnamefont{M.}~\bibnamefont{Onoda}},
  \bibinfo{author}{\bibfnamefont{T.}~\bibnamefont{Mizusaki}}, \bibnamefont{and}
  \bibinfo{author}{\bibfnamefont{H.}~\bibnamefont{Aoki}},
  \bibinfo{journal}{Phys.\ Rev. \ B} {\bibinfo{volume}{64}} (\bibinfo{year}{2001})
  \bibinfo{pages}{235315}.

\bibitem[{\citenamefont{Kallin and Halperin}(1984)}]{kallin}
\bibinfo{author}{\bibfnamefont{C.}~\bibnamefont{Kallin}} \bibnamefont{and}
  \bibinfo{author}{\bibfnamefont{B.~I.} \bibnamefont{Halperin}},
  \bibinfo{journal}{Phys. \ Rev. \ B} {\bibinfo{volume}{30}} (\bibinfo{year}{1984})
  \bibinfo{pages}{5655}.

\bibitem[{\citenamefont{Haldane and Rezayi}(1985)}]{haldane}
\bibinfo{author}{\bibfnamefont{F.~D.~M.} \bibnamefont{Haldane}}
  \bibnamefont{and} \bibinfo{author}{\bibfnamefont{E.~H.}
  \bibnamefont{Rezayi}}, \bibinfo{journal}{Phys.\ Rev. \ Lett.}
  {\bibinfo{volume}{54}} (\bibinfo{year}{1985}) \bibinfo{pages}{237}.

\bibitem[{\citenamefont{Girvin et~al.}(1985)\citenamefont{Girvin, MacDonald,
  and Platzman}}]{girvin}
\bibinfo{author}{\bibfnamefont{S.~M.} \bibnamefont{Girvin}},
  \bibinfo{author}{\bibfnamefont{A.~H.} \bibnamefont{MacDonald}},
  \bibnamefont{and} \bibinfo{author}{\bibfnamefont{P.~M.}
  \bibnamefont{Platzman}}, \bibinfo{journal}{Phys.\ Rev. \ Lett.}
  {\bibinfo{volume}{54}} (\bibinfo{year}{1985}) \bibinfo{pages}{581}.

\bibitem[{\citenamefont{Pinczuk et~al.}(1993)\citenamefont{Pinczuk, Dennis,
  Pfeiffer, and West}}]{pinc93}
\bibinfo{author}{\bibfnamefont{A.}~\bibnamefont{Pinczuk}},
  \bibinfo{author}{\bibfnamefont{B.~S.} \bibnamefont{Dennis}},
  \bibinfo{author}{\bibfnamefont{L.~N.} \bibnamefont{Pfeiffer}},
  \bibnamefont{and} \bibinfo{author}{\bibfnamefont{K.~W.} \bibnamefont{West}},
  \bibinfo{journal}{Phys.\ Rev. \ Lett.} {\bibinfo{volume}{70}} (\bibinfo{year}{1993})
  \bibinfo{pages}{3983}.

\bibitem[{\citenamefont{Pinczuk et~al.}(1995)\citenamefont{Pinczuk, Sohn,
  Dennis, Pfeiffer, and West}}]{pinc95}
\bibinfo{author}{\bibfnamefont{A.}~\bibnamefont{Pinczuk}},
  \bibinfo{author}{\bibfnamefont{L.~L.} \bibnamefont{Sohn}},
  \bibinfo{author}{\bibfnamefont{B.~S.} \bibnamefont{Dennis}},
  \bibinfo{author}{\bibfnamefont{L.~N.} \bibnamefont{Pfeiffer}},
  \bibnamefont{and} \bibinfo{author}{\bibfnamefont{K.~W.} \bibnamefont{West}},
  \bibinfo{journal}{Bull. \ Am. \ Phys. \ Soc.} {\bibinfo{volume}{40}} (\bibinfo{year}{1995})
  \bibinfo{pages}{515}.

\bibitem[{\citenamefont{Davies et~al.}(1997)\citenamefont{Davies, Harris, Ryan,
  and Turberfield}}]{davies97}
\bibinfo{author}{\bibfnamefont{H.~D.~M.} \bibnamefont{Davies}},
  \bibinfo{author}{\bibfnamefont{J.~C.} \bibnamefont{Harris}},
  \bibinfo{author}{\bibfnamefont{J.~F.} \bibnamefont{Ryan}}, \bibnamefont{and}
  \bibinfo{author}{\bibfnamefont{A.~J.} \bibnamefont{Turberfield}},
  \bibinfo{journal}{Phys.\ Rev. \ Lett.} {\bibinfo{volume}{78}} (\bibinfo{year}{1997})
  \bibinfo{pages}{4095}.

\bibitem[{\citenamefont{Sarma and Pinczuk}(1997)}]{perspectives}
\bibinfo{editor}{\bibfnamefont{S.~D.} \bibnamefont{Sarma}} \bibnamefont{and}
  \bibinfo{editor}{\bibfnamefont{A.}~\bibnamefont{Pinczuk}}, eds.,
  \emph{\bibinfo{title}{Perspectives in Quantum Hall Effects}}
  (\bibinfo{publisher}{Wiley, New York}, \bibinfo{year}{1997}).

\bibitem[{\citenamefont{Harries et~al.}(1998)\citenamefont{Harries, Davies,
  Ryan, and Turberfield}}]{davies99}
\bibinfo{author}{\bibfnamefont{J.~C.} \bibnamefont{Harries}},
  \bibinfo{author}{\bibfnamefont{H.~D.~M.} \bibnamefont{Davies}},
  \bibinfo{author}{\bibfnamefont{J.~F.} \bibnamefont{Ryan}}, \bibnamefont{and}
  \bibinfo{author}{\bibfnamefont{A.~J.} \bibnamefont{Turberfield}},
  \bibinfo{journal}{Physica B} {\bibinfo{volume}{258}} (\bibinfo{year}{1998})
  \bibinfo{pages}{44}.

\bibitem[{\citenamefont{Kang et~al.}(2000)\citenamefont{Kang, Pinczuk, Dennis,
  Eriksson, Pfeiffer, and West}}]{moonsoo00}
\bibinfo{author}{\bibfnamefont{M.}~\bibnamefont{Kang}},
  \bibinfo{author}{\bibfnamefont{A.}~\bibnamefont{Pinczuk}},
  \bibinfo{author}{\bibfnamefont{B.~S.} \bibnamefont{Dennis}},
  \bibinfo{author}{\bibfnamefont{M.~A.} \bibnamefont{Eriksson}},
  \bibinfo{author}{\bibfnamefont{L.~N.} \bibnamefont{Pfeiffer}},
  \bibnamefont{and} \bibinfo{author}{\bibfnamefont{K.~W.} \bibnamefont{West}},
  \bibinfo{journal}{Phys.\ Rev. \ Lett.} {\bibinfo{volume}{84}} (\bibinfo{year}{2000})
  \bibinfo{pages}{546}.

\bibitem[{\citenamefont{Kang et~al.}(2001)\citenamefont{Kang, Pinczuk, Dennis,
  Pfeiffer, and West}}]{moonsoo01}
\bibinfo{author}{\bibfnamefont{M.}~\bibnamefont{Kang}},
  \bibinfo{author}{\bibfnamefont{A.}~\bibnamefont{Pinczuk}},
  \bibinfo{author}{\bibfnamefont{B.~S.} \bibnamefont{Dennis}},
  \bibinfo{author}{\bibfnamefont{L.~N.} \bibnamefont{Pfeiffer}},
  \bibnamefont{and} \bibinfo{author}{\bibfnamefont{K.~W.} \bibnamefont{West}},
  \bibinfo{journal}{Phys.\ Rev. \ Lett.} {\bibinfo{volume}{86}} (\bibinfo{year}{2001})
  \bibinfo{pages}{2637}.

\bibitem[{\citenamefont{Scarola et~al.}(2000)\citenamefont{Scarola, Park, and
  Jain}}]{scar00}
\bibinfo{author}{\bibfnamefont{V.~W.} \bibnamefont{Scarola}},
  \bibinfo{author}{\bibfnamefont{K.}~\bibnamefont{Park}}, \bibnamefont{and}
  \bibinfo{author}{\bibfnamefont{J.~K.} \bibnamefont{Jain}},
  \bibinfo{journal}{Phys. \ Rev. \ B} {\bibinfo{volume}{61}} (\bibinfo{year}{2000})
  \bibinfo{pages}{13064}.

\bibitem[{\citenamefont{Dujovne et~al.}(2002)\citenamefont{Dujovne, Pinczuk,
  Kang, Dennis, Pfeiffer, and West}}]{dujovne}
\bibinfo{author}{\bibfnamefont{I.}~\bibnamefont{Dujovne}},
  \bibinfo{author}{\bibfnamefont{A.}~\bibnamefont{Pinczuk}},
  \bibinfo{author}{\bibfnamefont{M.}~\bibnamefont{Kang}},
  \bibinfo{author}{\bibfnamefont{B.~S.} \bibnamefont{Dennis}},
  \bibinfo{author}{\bibfnamefont{L.~N.} \bibnamefont{Pfeiffer}},
  \bibnamefont{and} \bibinfo{author}{\bibfnamefont{K.~W.} \bibnamefont{West}},
  \bibinfo{journal}{Phys.\ Rev. \ Lett.} {\bibinfo{volume}{90}} (\bibinfo{year}{2003})
  \bibinfo{pages}{036803}.

\bibitem[{\citenamefont{Dujovne et~al.}(2003)\citenamefont{Dujovne, Pinczuk,
  Kang, Dennis, Pfeiffer, and West}}]{dujovne03}
\bibinfo{author}{\bibfnamefont{I.}~\bibnamefont{Dujovne}},
  \bibinfo{author}{\bibfnamefont{A.}~\bibnamefont{Pinczuk}},
  \bibinfo{author}{\bibfnamefont{M.}~\bibnamefont{Kang}},
  \bibinfo{author}{\bibfnamefont{B.~S.} \bibnamefont{Dennis}},
  \bibinfo{author}{\bibfnamefont{L.~N.} \bibnamefont{Pfeiffer}},
  \bibnamefont{and} \bibinfo{author}{\bibfnamefont{K.~W.} \bibnamefont{West}},
  \bibinfo{journal}{work in progress}  (\bibinfo{year}{2003}).

\bibitem[{\citenamefont{Chakraborty}(2000)}]{tapash}
\bibinfo{author}{\bibfnamefont{T.}~\bibnamefont{Chakraborty}},
  \bibinfo{journal}{Adv. \ Phys.} {\bibinfo{volume}{49}} (\bibinfo{year}{2000})
  \bibinfo{pages}{959}.

\bibitem[{\citenamefont{Longo and Kallin}(1993)}]{longo}
\bibinfo{author}{\bibfnamefont{J.}~\bibnamefont{Longo}} \bibnamefont{and}
  \bibinfo{author}{\bibfnamefont{C.}~\bibnamefont{Kallin}},
  \bibinfo{journal}{Phys. \ Rev. \ B} {\bibinfo{volume}{47}} (\bibinfo{year}{1993}) \bibinfo{pages}{4429}.

\bibitem[{\citenamefont{Nakajima and Aoki}(1994)}]{aoki}
\bibinfo{author}{\bibfnamefont{T.}~\bibnamefont{Nakajima}} \bibnamefont{and}
  \bibinfo{author}{\bibfnamefont{H.}~\bibnamefont{Aoki}},
  \bibinfo{journal}{Phys. Rev. Lett.} {\bibinfo{volume}{73}} (\bibinfo{year}{1994})
  \bibinfo{pages}{3568}.

\bibitem[{\citenamefont{Lopez and Fradkin}(1995)}]{lopez95}
\bibinfo{author}{\bibfnamefont{A.}~\bibnamefont{Lopez}} \bibnamefont{and}
  \bibinfo{author}{\bibfnamefont{E.}~\bibnamefont{Fradkin}},
  \bibinfo{journal}{Phys.\ Rev. \ B} {\bibinfo{volume}{51}} (\bibinfo{year}{1995})
  \bibinfo{pages}{4347}.

\bibitem[{\citenamefont{Mandal and
  Jain}(2001{\natexlab{b}})}]{mandalactivation}
\bibinfo{author}{\bibfnamefont{S.~S.} \bibnamefont{Mandal}} \bibnamefont{and}
  \bibinfo{author}{\bibfnamefont{J.~K.} \bibnamefont{Jain}},
  \bibinfo{journal}{Phys.\ Rev. B} {\bibinfo{volume}{64}} (\bibinfo{year}{2001}{\natexlab{b}})
  \bibinfo{pages}{125310}.

\bibitem[{\citenamefont{Pan et~al.}(2003)\citenamefont{Pan, Stormer, Tsui,
  Pfeiffer, Baldwin, and West}}]{pan03}
\bibinfo{author}{\bibfnamefont{W.}~\bibnamefont{Pan}},
  \bibinfo{author}{\bibfnamefont{H.~L.} \bibnamefont{Stormer}},
  \bibinfo{author}{\bibfnamefont{D.~C.} \bibnamefont{Tsui}},
  \bibinfo{author}{\bibfnamefont{L.~N.} \bibnamefont{Pfeiffer}},
  \bibinfo{author}{\bibfnamefont{K.~W.} \bibnamefont{Baldwin}},
  \bibnamefont{and} \bibinfo{author}{\bibfnamefont{K.~W.} \bibnamefont{West}},
  \bibinfo{journal}{Phys. \ Rev. \ Lett.} {\bibinfo{volume}{90}} (\bibinfo{year}{2003})
  \bibinfo{pages}{016801}.

\bibitem[{\citenamefont{Scarola et~al.}(2002)\citenamefont{Scarola, Jain, and
  Rezayi}}]{scar02}
\bibinfo{author}{\bibfnamefont{V.~W.} \bibnamefont{Scarola}},
  \bibinfo{author}{\bibfnamefont{J.~K.} \bibnamefont{Jain}}, \bibnamefont{and}
  \bibinfo{author}{\bibfnamefont{E.~H.} \bibnamefont{Rezayi}},
  \bibinfo{journal}{Phys.\ Rev. \ Lett.} {\bibinfo{volume}{88}} (\bibinfo{year}{2002})
  \bibinfo{pages}{216804}.

\bibitem[{\citenamefont{Szlufarska et~al.}(2002)\citenamefont{Szlufarska,
  W\'{o}js, and Quinn}}]{izab02}
\bibinfo{author}{\bibfnamefont{I.}~\bibnamefont{Szlufarska}},
  \bibinfo{author}{\bibfnamefont{A.}~\bibnamefont{W\'{o}js}}, \bibnamefont{and}
  \bibinfo{author}{\bibfnamefont{J.~J.} \bibnamefont{Quinn}},
  \bibinfo{journal}{Physica E} {\bibinfo{volume}{12}} (\bibinfo{year}{2002})
  \bibinfo{pages}{59}.

\bibitem[{\citenamefont{Pinczuk et~al.}(1988)\citenamefont{Pinczuk, Valladares,
  Heiman, Gossard, English, Tu, Pfeiffer, and West}}]{pinc88}
\bibinfo{author}{\bibfnamefont{A.}~\bibnamefont{Pinczuk}},
  \bibinfo{author}{\bibfnamefont{J.~P.} \bibnamefont{Valladares}},
  \bibinfo{author}{\bibfnamefont{D.}~\bibnamefont{Heiman}},
  \bibinfo{author}{\bibfnamefont{A.~C.} \bibnamefont{Gossard}},
  \bibinfo{author}{\bibfnamefont{J.~H.} \bibnamefont{English}},
  \bibinfo{author}{\bibfnamefont{C.~W.} \bibnamefont{Tu}},
  \bibinfo{author}{\bibfnamefont{L.~N.} \bibnamefont{Pfeiffer}},
  \bibnamefont{and} \bibinfo{author}{\bibfnamefont{K.~W.} \bibnamefont{West}},
  \bibinfo{journal}{Phys.\ Rev. \ Lett.} {\bibinfo{volume}{61}} (\bibinfo{year}{1988})
  \bibinfo{pages}{2701}.

\bibitem[{\citenamefont{Marmorkos and Sarma}(1992)}]{marmorkos}
\bibinfo{author}{\bibfnamefont{I.~K.} \bibnamefont{Marmorkos}}
  \bibnamefont{and} \bibinfo{author}{\bibfnamefont{S.~D.} \bibnamefont{Sarma}},
  \bibinfo{journal}{Phys.\ Rev. \ B} {\bibinfo{volume}{45}} (\bibinfo{year}{1992})
  \bibinfo{pages}{13396}.

\bibitem[{\citenamefont{Kang et~al.}(1997)\citenamefont{Kang, Young, hannahs,
  Palm, Campman, and Gossard}}]{wkang}
\bibinfo{author}{\bibfnamefont{W.}~\bibnamefont{Kang}},
  \bibinfo{author}{\bibfnamefont{J.~B.} \bibnamefont{Young}},
  \bibinfo{author}{\bibfnamefont{S.~T.} \bibnamefont{Hannahs}},
  \bibinfo{author}{\bibfnamefont{E.}~\bibnamefont{Palm}},
  \bibinfo{author}{\bibfnamefont{K.~L.} \bibnamefont{Campman}},
  \bibnamefont{and} \bibinfo{author}{\bibfnamefont{A.~C.}
  \bibnamefont{Gossard}}, \bibinfo{journal}{Phys. \ Rev. B}
  {\bibinfo{volume}{56}} (\bibinfo{year}{1997}) \bibinfo{pages}{12776}.

\bibitem[{\citenamefont{Kukushkin et~al.}(1999)\citenamefont{Kukushkin,
  v.~Klitzing, and Eberl}}]{kukushkin}
\bibinfo{author}{\bibfnamefont{I.~V.} \bibnamefont{Kukushkin}},
  \bibinfo{author}{\bibfnamefont{K.}~\bibnamefont{v.~Klitzing}},
  \bibnamefont{and} \bibinfo{author}{\bibfnamefont{K.}~\bibnamefont{Eberl}},
  \bibinfo{journal}{Phys.\ Rev. \ Lett.} {\bibinfo{volume}{82}} (\bibinfo{year}{1999})
  \bibinfo{pages}{3665}.

\bibitem[{\citenamefont{Pinczuk et~al.}(1992)\citenamefont{Pinczuk, Dennis,
  Heiman, Kallin, Brey, Tejedor, Schmitt-Rink, Pfeiffer, and West}}]{pinc92}
\bibinfo{author}{\bibfnamefont{A.}~\bibnamefont{Pinczuk}},
  \bibinfo{author}{\bibfnamefont{B.~S.} \bibnamefont{Dennis}},
  \bibinfo{author}{\bibfnamefont{D.}~\bibnamefont{Heiman}},
  \bibinfo{author}{\bibfnamefont{C.}~\bibnamefont{Kallin}},
  \bibinfo{author}{\bibfnamefont{L.}~\bibnamefont{Brey}},
  \bibinfo{author}{\bibfnamefont{C.}~\bibnamefont{Tejedor}},
  \bibinfo{author}{\bibfnamefont{S.}~\bibnamefont{Schmitt-Rink}},
  \bibinfo{author}{\bibfnamefont{L.~N.} \bibnamefont{Pfeiffer}},
  \bibnamefont{and} \bibinfo{author}{\bibfnamefont{K.~W.} \bibnamefont{West}},
  \bibinfo{journal}{Phys. Rev. Lett.} {\bibinfo{volume}{68}} (\bibinfo{year}{1992})
  \bibinfo{pages}{3623}.

\bibitem[{\citenamefont{Pan et~al.}(2002)\citenamefont{Pan, Stormer, Tsui,
  Pfeiffer, Baldwin, and West}}]{pan}
\bibinfo{author}{\bibfnamefont{W.}~\bibnamefont{Pan}},
  \bibinfo{author}{\bibfnamefont{H.~L.} \bibnamefont{Stormer}},
  \bibinfo{author}{\bibfnamefont{D.~C.} \bibnamefont{Tsui}},
  \bibinfo{author}{\bibfnamefont{L.~N.} \bibnamefont{Pfeiffer}},
  \bibinfo{author}{\bibfnamefont{K.~W.} \bibnamefont{Baldwin}},
  \bibnamefont{and} \bibinfo{author}{\bibfnamefont{K.~W.} \bibnamefont{West}},
  \bibinfo{journal}{Phys. \ Rev. \ Lett.} {\bibinfo{volume}{88}} (\bibinfo{year}{2002})
  \bibinfo{pages}{176802}.

\bibitem[{\citenamefont{Szlufarska et~al.}(2001)\citenamefont{Szlufarska,
  W\'{o}js, and Quinn}}]{izab}
\bibinfo{author}{\bibfnamefont{I.}~\bibnamefont{Szlufarska}},
  \bibinfo{author}{\bibfnamefont{A.}~\bibnamefont{W\'{o}js}}, \bibnamefont{and}
  \bibinfo{author}{\bibfnamefont{J.~J.} \bibnamefont{Quinn}},
  \bibinfo{journal}{Phys. \ Rev. \ B} {\bibinfo{volume}{64}} (\bibinfo{year}{2001})
  \bibinfo{pages}{165318}.

\end{thebibliography}

\end{document}